\documentclass[11pt]{article}

\usepackage{amsmath,amssymb,amsbsy,amstext,mathrsfs}
\usepackage{caption}

\newcommand{\rom}[1]{\mathrm{#1}}
\newcommand{\tf}{\tilde{f}}
\newcommand{\tldmu}{\tilde{\mu}}

\usepackage{amsmath,amssymb,graphicx}
\usepackage{hyperref}
\usepackage{tikz}
\usetikzlibrary{decorations.pathmorphing}

\usepackage{multicol,color,longtable}

\definecolor{darkred}{rgb}{0.65,0.15,0}
\hypersetup{pdfborder={0 0 0},colorlinks=true,urlcolor=darkred,citecolor=blue,linkcolor=darkred,linktocpage=true}

\definecolor{AV}{rgb}{0.65,0.0,0}
\definecolor{AK}{rgb}{0,0,1}
\definecolor{DK}{rgb}{0.6,0.4,0}

\newcommand{\mathM}{\mathcal{M}}

\newcommand\be{\begin{equation}}
\newcommand\ee{\end{equation}}
\def\bea{\begin{eqnarray}}
\def\eea{\end{eqnarray}}

\newcommand{\beq}{\begin{eqnarray}}
\newcommand{\eeq}{\end{eqnarray}}

\makeatletter

\@addtoreset{equation}{section}
\makeatother

\begin{document}

\thispagestyle{empty}

\begin{center}
{\Large \bf Charged Vaidya Solution Satisfies  Weak  Energy Condition}\\[10mm]

\vspace{8mm}
\normalsize
{\large   Soumyabrata Chatterjee, Suman Ganguli,
and
Amitabh Virmani}

\vspace{10mm}
{Institute of Physics\\
Sachivalaya Marg, Bhubaneshwar, Odisha, India 751005}
\vspace{15mm}
\hrule
\vspace{10mm}

\begin{tabular}{p{12cm}}
{\small
The external matter stress-tensor supporting charged Vaidya solution appears to violate weak energy condition in certain region of the spacetime. Motivated by this, a new interpretation of charged Vaidya solution was proposed by Ori \cite{Ori} in which the energy condition continues to be satisfied. In this construction, one glues an outgoing Vaidya solution to the original ingoing Vaidya solution  provided the  surface where the external stress-tensor vanishes is spacelike. We revisit this study and extend it to higher-dimensions, to AdS settings, and to higher-derivative $f(R)$ theories. In asymptotically flat space context, we explore in detail the case when the mass function $m(v)$ is   proportional to the charge function $q(v)$. When  the proportionality constant $\nu = q(v)/m(v)$ lies in between zero and one, we show that the  surface  where the external stress-tensor vanishes is spacelike and lies in between the inner and  outer apparent horizons.
}
\end{tabular}
\vspace{10mm}
\hrule
\end{center}

\newpage
\setcounter{page}{1}

\tableofcontents

\section{Introduction}
Evolution of charged null shells in spherically symmetric settings has been a source of confusion in the past \cite{Bonnor:1970zz, LakeZannias, SullivanIsrael, Kaminaga:1988pg}. Despite the simple form of the charged Vaidya solution and its seemingly simple interpretation, it suffers from a rather serious conceptual difficulty: it seems to violate the weak (as well as null) energy condition in certain region of the spacetime. In four-dimensional asymptotically flat context the region where weak energy condition is violated is,
\be
r < r_c(v) = \frac{q(v) \dot q(v)}{\dot m(v)},
\label{region}
\ee
where $m(v)$ and $q(v)$ are respectively the mass and charge functions of the ingoing charged Vaidya solution, and $v$ is the advanced null coordinate.  In modern literature this feature has been argued to be related to the possible violation of  strong sub-additivity of holographic entanglement entropy \cite{Caceres:2013dma}.

For the ingoing charged Vaidya solution, the spherical shells of ``charged photons'' all start at infinity and flow inwards. The key physical question is therefore whether these shells enter the region \eqref{region} where the energy conditions can be violated. Ori \cite{Ori} in 1991 did a systematic and thorough analysis of this question. In this paper we revisit that study and extend it in a number of ways.

The key confusion in the earlier literature was the assumption that  the  Lorentz force does not matter and the charged null fluid flows along the radial null \emph{geodesics}.
 In the Reissner-Nordstr\"om geometry the radial null geodesics go  all the way from $r=\infty$ to $r=0$. Assuming that the Lorentz force is irrelevant, earlier authors concluded that nothing can prevent charged null shells from entering the region where energy conditions are violated. Ori, on the other hand, showed that the assumption that the Lorentz force can be ignored is incorrect (in fact inconsistent, as we review below). By properly taking it into account  he showed that the weak energy condition continues to be satisfied.

The most important outcome of this rather intricate analysis was as follows: locally the Lorentz orbits are tangent to \emph{one} of the two (ingoing or outgoing) radial null geodesics, however, at $r=r_c$ the tangent direction flips. The null vector $k^a$ along the Lorentz orbit vanishes at $r=r_c$. The physical continuation of the initially ingoing orbit at $r=r_c$ is to switch to the outgoing one.

Intuitive  explanation of this result is rather straightforward. Charged particle moving in charged background looses its kinetic energy due to electromagnetic repulsion at it falls inwards. At $r=r_c$ its kinetic energy is reduced to zero. At this point the particle starts moving outwards.

Precisely the same  intuitive picture applies to the charged Vaidya spacetime. At $r=r_c$ the charged null shells switch from ingoing to outgoing orbits, and there is no violation of the energy conditions. In this paper, after providing a concise summary of Ori's work, we extend the analysis in the following ways:
\begin{enumerate}
\item We adapt Ori's arguments to the AdS and higher-dimensional settings. This extension brings in some minor technical changes in the details; the physical picture remains unchanged.
\item We adapt the same set of arguments to  $f(R)$ theories coupled to a Maxwell field. Again this extension work straightforwardly. It suggests that the picture suggested by Ori is valid in other settings as well, e.g., Einstein-Gauss-Bonnet gravity coupled to a Maxwell field.
\item In the asymptotically flat context, we analyse in some detail the case when the mass function $m(v)$ is  proportional to the charge function $q(v)$. For the physically interesting case where the proportionality constant $\nu = q(v)/m(v)$ is in between zero and one, we show that the  surface  where the external stress-tensor vanishes is spacelike and lies in between the inner and the outer apparent horizons. Since this surface is spacelike, to its future is an \emph{outgoing} Vaidya solution. We discuss this gluing construction in some detail.
\end{enumerate}

The rest of the paper is organised as follows. In section \ref{sec:prelims} we start with a review of the charged Vaidya solution and its standard interpretation. We review the fact that the interpretation of charged null fluid as made up of discrete charged massless particles is inconsistent with the geodesic motion assumption.  We highlight how the null Lorentz force equation appears from such an analysis.

In section \ref{sec:probe} we analyse radial orbits of massless
charged particles in higher-dimensional AdS Reissner-Nordstr\"om background. In this analysis we focus our attention on the behaviour of orbits near the vanishing point $k^a = 0$. This probe analysis already shows how initially ingoing orbits switch to outgoing ones. We also analyse the behaviour of orbits near a generic vanishing point.

In section \ref{sec:CVS} we return to the interpretation of the charged Vaidya solution and apply the results of the former sections. The evolution beyond the hypersurface
of the vanishing points is discussed. $f(R)$-Maxwell set-up is discussed in section \ref{sec:fR}. We end with a brief summary of our results and some speculative discussion in section \ref{sec:disc}.  Power law profiles for $m(v)$ and $q(v)$ functions in  asymptotically flat context are further discussed in appendix \ref{app:power}.

\section{Preliminaries: Charged Vaidya solution and the Lorentz force equation}
\label{sec:prelims}
In  section \ref{prelim1} we start with a brief review of the charged Vaidya solution and its standard interpretation. Then, following Ori \cite{Ori},  we show in section \ref{prelim2}   that this  interpretation is inconsistent with the physical picture that the shell is made up of discrete charged null particles. Consistency of the equations of motion requires us to replace the geodesic equation used in the standard interpretation with the null Lorentz force equation.
\subsection{Charged Vaidya solution}
\label{prelim1}
Let us consider the charged Vaidya solution for the
following Lagrangian in $d$ spacetime dimensions,
\be
L = \left(\frac{1}{16 \pi G} \right) \sqrt{-g}\left[R - 2 \Lambda - \frac{1}{4}F_{ab}F^{ab}\right] + L_\mathrm{matter},
\ee
where $L_\mathrm{matter}$ is the Lagrangian for the external matter source. The equations of motion are
\be
R_{ab} - \frac{1}{2} R g_{ab}  + \Lambda g_{ab}  = 8 \pi \left( T_{ab}^{\rom{(em)}}  + T^{\rom{(m)}}_{ab} \right).
\ee
where
\be
 T_{ab}^{\rom{(em)}}  =  \frac{1}{16 \pi} \left( F_{ac}F_{b}{}^{c} - \frac{1}{4}g_{ab}F_{cd}F^{cd} \right),
 \label{TEM}
\ee
and $T^{\rom{(m)}}_{ab}$ is the stress-tensor of the external matter source. The cosmological constant parameter $\Lambda$ is related to a characteristic length scale $l$ as $\Lambda = - \frac{(d-1) (d-2)}{2l^2}$. The length scale $l$ is a more convenient quantity to work with. We take the cosmological constant term to be negative; we are only interested in asymptotically anti-de-Sitter and flat (obtained by sending $l \to \infty$)
settings. The ingoing charged Vaidya metric is described by the line element
\be
ds^2 = - f(r,v) dv^2 + 2 dr dv + r^2 d \Omega_{d-2}^2,
\label{line_element}
\ee
where the metric function $f(r,v)$ is
\be
f(r,v) = 1 - \frac{2 m(v)}{r^{d-3}}  + \frac{q^2(v)}{r^{2(d-3)}} + \frac{r^2}{l^2},
\ee
and $d \Omega_{d-2}^2$  is the line element on the  unit round $(d-2)$ sphere.
The Maxwell potential $A_a$ supporting the solution is,
\be
A_v =  -\sqrt{\frac{2(d-2)}{d-3}} \frac{q(v)}{r^{d-3}}.
\label{vector}
\ee
 %The parameter $\epsilon$ in \eqref{line_element} distinguishes between  ingoing and  outgoing solutions: $\epsilon = + 1$ corresponds to metric for ingoing charged null fluid and $\epsilon = -1$ to the outgoing charged null fluid. For the most part we are interested in the ingoing solution; unless otherwise mentioned from now onwards we take $\epsilon = + 1$.

The external matter stress tensor needed to support this solution is
\be
8 \pi T_{vv}= \frac{d-2}{r^{2d-5}} \left(r^{d-3}\dot m - q \dot q \right). \label{external_matterST}
\ee
In particular, it is of the form
\be
T_{ab} = \rho k_a k_b,\label{matterST}
\ee
where the vector $k^a$ is
(ingoing) null vector tangent to the  $v= \ $constant lines. Normalisations of $\rho$ and the null vector $k^a$ are not fixed by the form \eqref{matterST}. Further physical inputs are required to fix these normalisations. In the standard interpretation, which has been a source of confusion in the past, the choice
\be
k_{a} = - \partial_a v = - \delta^v_a,
\ee
is made. With this choice $k^{a}$ is the null vector generating the ingoing null  \emph{geodesics}. Therefore, the charged null fluid is thought of as flowing along these null  geodesics. Under this assumption, charged null particles starting from $r=\infty$  run all the way to $r=0$ along $v= \  $constant lines.

\paragraph{Null energy condition:}
Generators of radial null orbits $ l^a = \left(l^v,l^r, 0 ,\cdots, 0 \right)$ that intersect the null fluid satisfies
$ f(r,v)l^v =  2 l^r.$ The local energy density  ${\mathcal K}= T_{ab}  l^a l^b$ ``experienced" by these trajectories intersecting the charged null fluid simplifies to,
\be
\label{NEC}
{\mathcal K}(v,r)  = \left( \frac{d-2}{8 \pi} \right) \frac{1}{ r^{2d-5}} \left( r^{d-3}  - \frac{q  \dot{q}}{\dot{m}}   \right)  \dot{m} l_r^2.
\ee
Clearly, for
\be
r^{d-3}   <  r_c^{d-3}  := \frac{q  \dot{q}}{\dot{m}}, \label{bad_region}
\ee
${\mathcal K}(v,r)$ becomes negative; the null energy condition is violated.

\paragraph{Weak energy condition:}  In a similar manner the energy density
${\mathcal E} = T_{ab} u^a u^b$ measured by a timelike observer with normalised velocity $u^a = \left(u^v, u^r, 0 ,\cdots,0\right)$ simplifies to,
\be
\label{WEC}
{\mathcal E}(v,r)  = \frac{(d-2)(d-3)}{16 \pi} \frac{q^2}{r^{2(d-2)}} + \left( \frac{d-2}{8 \pi} \right)\frac{1}{ r^{2d-5}} \left( r^{d-3}  - \frac{q\dot{q}}{\dot{m}}   \right)  \dot{m}  u_r^2.
\ee
The first term  corresponds to the ``Maxwellian'' energy density due to the presence of electric charge. The second term is the energy density of the particles constituting the collapsing charged null shell. Once again, in the region  \eqref{bad_region}  the second term becomes negative. Moreover, for sufficiently large $u_r$ this term dominates, and the total ${\mathcal E}(v,r)$ becomes negative; weak energy condition is violated.

\subsection{Lorentz force equation}
\label{prelim2}
Now we demonstrate that the geodesic motion assumption is inconsistent. Above we distinguished between the matter and electromagnetic stress tensor. Let us look at the
equation arising from Bianchi identity, i.e., the conservation of the total stress-tensor. We get
\be
8 \pi \nabla^aT^{\rom{(m)}}_{ab} =  -\frac{1}{2} (\nabla^a F_{ac})F_{b}{}^{c} - \frac{1}{2} F_{ac} (\nabla^a
F_{b}{}^{c}) + \frac{1}{4} F_{cd} \nabla_b F^{cd}.
\ee
This expression upon using the Maxwell equation
\be
\nabla^a F_{ac} = -16 \pi J^{(e)}_c \label{Maxwell_eq}
\ee  and the Bianchi identity for the Maxwell field becomes
%\be
% \nabla^a T^{\rom{(m)}}_{ab} = \frac{1}{2} F^{bc} J^{(e)}_c - \frac{1}{2} F_{ac} (\nabla^a F^{bc}) +
%\frac{1}{4}F_{cd} \nabla^b F^{cd}.
%\ee
%We can rewrite it as
%\be
%\nabla^a T^{\rom{(m)}}_{ab} = \frac{1}{2} F^{bc} J^{(e)}_c +  \frac{1}{4} F_{cd} (\nabla^c F^{db} + \nabla^d
%F^{bc}+ \nabla^b F^{cd}).
%\ee
%Using Bianchi identity for the Maxwell field we get
\be
 \nabla^a T^{\rom{(m)}}_{ab} = F^{bc} J^{(e)}_c. \label{Maxwell_eq2}
\ee
Now recalling that the matter stress-tensor is of the form \eqref{matterST}, equation \eqref{Maxwell_eq2} becomes,
\be
\nabla^a(\rho k_a k_b) = F^{bc} J^{(e)}_c. \label{Lorentz}
\ee

We adopt the hydrodynamic point of view and think
of the charged null fluid as a stream of discrete charged massless particles. We associate a
conserved null current associated with the flow
\begin{align}
J_a &= \rho k_a, & \nabla^a J_a &= 0. \label{continuity}
\end{align}
Similarly $J^{(e)}_a:= \rho_e k_a $ is thought of as conserved electric current; $ \rho_e$ being the electric charge density.
The ratio $e := J^{(e)}_a/J_a = \rho_e/\rho$ is the charge per particle. The conservation of
$e$ follows, and it reflects the fact that each particle carries its own
electric charge.
Using these inputs, equation \eqref{Lorentz} becomes
\be
 k^a \nabla_a k^b =  e F^{b}{}_{c} k^c. \label{Lorentz_main}
\ee

Equation \eqref{Lorentz_main} is the key equation for the consideration of this paper. It is not the geodesic equation. It is the massless analog of the Lorentz force equation for a massive particle; for simplicity we continue to call it the Lorentz force equation. We call the parameter $\lambda$ along the orbits satisfying  \eqref{Lorentz_main} the Lorentz  parameter. In general, the Lorentz  parameter is different from the affine parameter, however, when $F^{b}{}_{c} k^c = 0$, it becomes identical to the affine parameter. In that case the  Lorentz orbit becomes a null geodesic.

\section{Probe computation}
\label{sec:probe}

To understand the full implications of the Lorentz force equation \eqref{Lorentz_main}, it is instructive to start with a probe  computation.  Due to the electromagnetic repulsion the magnitude of $k^a$ for infalling
``charged photons'' gradually decreases and at some spacetime point $k^a$ vanishes. Our focus in this analysis is on the behaviour of orbits near the vanishing point $k^a = 0$.  In section \ref{sec:spherical_symmetry} we regard the higher-dimensional AdS Reissner-Nordstr\"om metric and the associated electromagnetic field as a given background, and study the Lorentz orbits in this background.
 This probe analysis already shows how initially ingoing orbits switch to outgoing ones. We  analyse the behaviour of orbits near a generic vanishing point in higher-dimensions in section \ref{sec:generic}.  Other studies of the motion of charged particles in the field of (rotating) charged black holes include \cite{Balek, Pugliese:2011py}.

\subsection{Vanishing point in the spherically symmetric context}
\label{sec:spherical_symmetry}

The AdS Reissner-Nordstr\"om background in Schwarzschild coordinates in $d$ dimensions is
\be
ds_d^2=-f  dt^2+f^{-1}dr^2+r^2 d\Omega_{d-2}^2,
\ee
where the metric function $f$ is
\be
f(r)= 1-\frac{2m}{r^{d-3}}+\frac{q^2}{r^{2(d-3)}} + \frac{r^2}{l^2}.
\ee
  The only non-vanishing components of field strength
tensor are
\be
F_{rt}=-F_{tr}=  \sqrt{2(d-2)(d-3)}\frac{q}{r^{d-2}}.
\ee

To solve the  Lorentz force equation \eqref{Lorentz_main}  for radial trajectories in this background  we take  $k_a$ of the form
$k_a=(k_t(r),k_r(r),0,\ldots ,0). $ With this ansatz equations can be readily solved.  For ingoing orbits we obtain
\bea
k_r &=& f^{-1}k_t, \qquad k^r  = k_t, \\
k_t &=& -\left(\kappa-\sqrt{\frac{2(d-2)}{d-3}}\frac{e q}{r^{d-3}}\right). \label{kdownt}
\eea
The parameter $\kappa$ is interpreted as the  energy of the infalling particle measured by a static observer at infinity.
The vanishing point is located at
\be
r^{d-3}= r_c^{d-3} : = \sqrt{\frac{2(d-2)}{d-3}}\frac{e q}{\kappa},
\label{rvan}
\ee
provided $eq > 0$, $\kappa > 0$.

We can parametrise such an orbit with a real parameter $\lambda$ such that $k^a=\frac{dx^a}{d\lambda}$.
We find by direct integration,
\begin{align}
\lambda &=  - \frac{r}{\kappa}- \frac{r_c}{\kappa} \log |r- r_c| + c,  & \mbox{for} \qquad d = 4,\\
\lambda &= -\frac{r}{\kappa} F\left(1,\frac{1}{3-d}, \frac{d-4}{d-3}, \frac{r_c^{d-3}}{r^{d-3}}\right)+c,  & \mbox{for} \qquad d > 4,
\end{align}
where $c$ is an integration constant. It is fixed by the initial conditions; its numerical value is not important for the arguments below. For simplicity we set $c = 0$. The function $F$ is the hypergeometric function. For $d=4$ as  $r$ approaches $r_c$ the Lorentz parameter $\lambda$ diverges to $+\infty$.
For $d > 4$ at $r=r_c$, the parameter $\lambda$ becomes
\be
\lambda=-\frac{r_c}{\kappa} F\left(1,\frac{1}{3-d}, \frac{d-4}{d-3}, 1\right),
\ee
which also diverges to $+ \infty$. This can be easily seen by the identity
\be
F\left(a,b,c,1\right)  = \frac{\Gamma(c)\Gamma(c-a-b)}{\Gamma(c-a)\Gamma(c-b)}.
\ee

At the vanishing point the momentum vector $k^a$ vanishes. How to continue the orbit beyond the vanishing point? To answer this question  further physical inputs are required. Mathematically there are two possibilities:
\begin{enumerate}
\item  \textbf{Non-bouncing or ingoing continuation:} This corresponds to $k^r=k_t$. To begin with $k_t < 0$, see equation \eqref{kdownt}. Across the vanishing point $k_t$ (and hence $k^r$) changes sign. Thus, for this case, \be \frac{dr}{dt}=\frac{k^r}{k^t}=-f, \ee preserves its sign as the orbit crosses the  vanishing point. As a result the orbit continues along the same (ingoing) null direction. That is why we call it the non-bouncing or the ingoing continuation.

It is instructive to analyse the non-bouncing continuation in two separate cases depending on whether the metric function $f$ is positive or negative at the vanishing point $r=r_c$.
\begin{enumerate}
\item $f > 0$: In this case $\frac{dr}{dt}<0$ to begin with at $r=r_c$. It continues to remain negative beyond the vanishing point. However, the null vector $k^a$ changes sign -- from future directed it becomes past directed -- so the Killing energy carried by the infalling particles changes sign.
\item $f < 0$: In this case $\frac{dr}{dt}>0$ to begin with at $r=r_c$. That is to say that for an ingoing particle in between the inner and the outer horizon, both $r$ and $t$ decrease. Beyond the vanishing point $\frac{dr}{dt}$ does not change sign. However, once again, the null vector $k^a$ changes sign so the Killing energy carried by the infalling particles changes sign.
\end{enumerate}

Since the null vector $k^a$ changes from future directed to past directed, this is our first hint that this continuation is not the physical one.

\item \textbf{Bouncing or outgoing continuation:}  This corresponds to  taking $k^r=-k_t$ beyond the vanishing point. For this case $\frac{dr}{dt}$
changes sign. The Lorentz orbit switches from ingoing to outgoing. For this reason we call this the bouncing or the outgoing continuation.

Once again it is instructive to analyse the bouncing continuation in two separate cases depending on whether the metric function $f$ is positive or negative at the vanishing point $r=r_c$.

\begin{enumerate}
\item $f > 0$: To begin with $k^r=k_t < 0$, and therefore $\frac{dr}{dt}=-f<0$, i.e.,  $r$ decreases with time.
Beyond the vanishing point,
\be
\frac{dr}{dt}= \frac{k^r}{k^t} = \frac{k^r}{g^{tt}k_t}=+f >0.
\ee
Therefore, $r$ starts to increase with time. The orbit switches  from the ingoing to  outgoing. From equation \eqref{kdownt} we note that  $k_t$ continues to be negative beyond the vanishing point.  Hence, $k^a$ continues to be future directed null vector; the Killing energy carried by
the particle continues to be positive.

\item $f < 0$:  In this situation the particle is in between the inner and the outer horizon. For future directed causal curves in this region the coordinate $r$ always decreases. To began with $\frac{dr}{dt} = - f $ is positive. Beyond the vanishing point $\frac{dr}{dt}$ is negative $\frac{dr}{dt} = f < 0$. To begin with  $k_t < 0$. From equation \eqref{kdownt} we note that  beyond the vanishing point, in the present case, $k_t$ changes sign, as the coordinate $r$ being timelike continues to decrease beyond $r_c$. However, $k^r$ does not change sign. As a result $k^a$ remains future directed.
\end{enumerate}
Since the null vector $k^a$ continues to be  future directed, the bouncing continuation seems to a  more physical one.

\end{enumerate}

 To see that the bouncing continuation is the physical continuation we introduce a small conserved angular momentum $k_\phi=L >0$. The introduction of the parameter $L$ prevents the vanishing of the vector $k^a$ and therefore removes the above mentioned mathematical ambiguity. In the limit $L \to 0$  these orbits converge to bouncing radial orbits, thus establishing that the bouncing continuation is indeed the physical one \cite{Ori}.

  Let us take $k_a=(k_t,k_r,0, \ldots, 0, L)$, and consider Lorentz null orbits in an appropriate equatorial plane where $g_{\phi \phi} = r^2$.  The null condition gives
\begin{equation}
 \left(k^r\right)^2=\left(k_t\right)^2-\frac{f}{r^2}L^2. \label{null_condition}
\end{equation}
The Lorentz force equation  \eqref{Lorentz_main} can also be written as
\be
\frac{dk^a}{d\lambda}  + \Gamma^a_{bc} k^b k^c = e F^{a}{}_{b} k^b, \label{lorentz_force_parameter}
\ee
where
$\Gamma^a_{bc}$ is the Christoffel symbol.  Once again the $k_t$ equation can be readily integrated with solution exactly the same expression as \eqref{kdownt}. The $k^r$ equation can be written as
\begin{equation}
 \frac{dk^r}{d\lambda}= \frac{1}{2} f^{-1} f'(k^r)^2 - \frac{1}{2} f f' (k^t)^2 + r f (k^\phi)^2 + \sqrt{2(d-2)(d-3)} \left(\frac{e q}{r^{d-2}} \right) f k^t. \label{radial}
\end{equation}
We are interested in the location of the turning point in the radial coordinate, i.e., the point where $k^r$ vanishes. In that case  equation \eqref{radial} simplifies to
\be
 \frac{dk^r}{d\lambda}= - \frac{1}{2} f f' (k^t)^2 + r f (k^\phi)^2 +\sqrt{2(d-2)(d-3)} \left(\frac{e q}{r^{d-2}} \right) f k^t. \label{radial_kr_zero}
\ee
Substituting $k^r=0$ in  \eqref{null_condition} gives
\be
(k_t)^2 = \frac{f L^2}{r^2}. \label{kr_equal_zero}
\ee
Once again it is useful to analyse  $f > 0$ and  $f < 0$ cases for the location of the turning point $k^r=0$ separately.
%%%%%%%%%
\begin{enumerate}
\item  $f>0$: Substituting  \eqref{kdownt} in   \eqref{kr_equal_zero} we note  that the turning point is located at\footnote{Recall that initially $k_t < 0$.}  the solution of the equation
\begin{equation}
 r^{d-3} = r_c^{d-3} + \frac{\sqrt{f} L}{\kappa} r^{d-4}.
\end{equation}
For small values of $L$, we immediately see that the solution of this equation is at $r > r_c$.  Moreover, for small $L$,  equation \eqref{radial_kr_zero} becomes
\be
 \frac{dk^r}{d\lambda}= +\sqrt{2(d-2)(d-3)} \left(\frac{e q}{r^{d-2}} \right) \frac{\sqrt{f}L}{r} + \mathcal{O}(L^2) > 0.
\ee
Therefore, the particle  bounces with positive $k^r$. Since the turning point is at $r > r_c$, $k_t$ does not vanish at the turning point. It continues to maintain its sign beyond the turning point; $k^a$ remains future directed.

Taking  $L \to 0$ limit in this analysis we conclude that the bouncing continuation of the radial null orbits is the physical one.

\item $f<0$:  In this case the particle is in between the inner and  outer horizon. For future directed causal curves in this region the coordinate $r$ always decreases. Therefore $r=r_c$ is reached with negative $k^r$. The null condition \eqref{null_condition} becomes
\begin{equation}
 \left(k^r\right)^2=\left(k_t\right)^2 + \frac{|f|}{r^2}L^2. \label{null_condition2}
\end{equation}
From equation \eqref{kdownt} we note that $k_t$ vanishes at $r=r_c$ and changes sign as $r$ continues to decrease. From \eqref{null_condition2} we see that  $k^r$ never vanishes.  It continues to be negative and hence future directed.

Once again, taking $L\rightarrow 0$  limit  in this analysis we conclude that the bouncing continuation of the radial null orbits is the physical one.

\end{enumerate}
To summarise: the above probe analysis allows us to conclude that the initially ingoing radial null Lorentz orbits switch to the outgoing ones.

\subsection{Properties of the generic vanishing point}
\label{sec:generic}
In this section we analyse the local behaviour of null Lorentz orbits near a generic vanishing point. We perform this study in an infinitesimal small neighbourhood of the vanishing point; spacetime curvature and  variations of the electromagnetic field are not important \cite{Ori}. We can obtain relevant properties of vanishing points and behaviour of orbits near them by working in flat spacetime with constant electromagnetic field. Equation \eqref{lorentz_force_parameter}  simplifies to the form
\be
\frac{dk^a}{d\lambda} = e F^{a}{}_{b} k^b.\label{lorentz_force_parameter_flat}
\ee

In Cartesian coordinates $(t,x^1,x^2,\ldots, x^{d-1})$ a canonical form of the electromagnetic field-strength tensor $F_{ab}$ in $d$ dimensions
\cite{Moita} is\footnote{There are also other canonical forms for $F_{ab}$ in higher dimensions \cite{Moita}, however, we restrict our study to this case only.}
\be
F_{ab} =
 \begin{pmatrix}
  0 & 0       & 0         &   0 &     \cdots & E \\
  0 & 0       & B   &   0 &     \cdots & 0 \\
  0 & -B  &0          &     0&    \cdots & 0 \\
  0 & 0  &0          &     0&    \cdots & 0 \\
  \vdots  & \vdots        &    \vdots &     \vdots & \ddots & \vdots\\
  -E & 0  &      \cdots         & \cdots  &      \cdots & 0
 \end{pmatrix}.
 \ee

For ease of writing we use the notation $x^1=x, \ x^2 =y $ and $x^{d-1} = z$. The equations of motion \eqref{lorentz_force_parameter_flat} are
\begin{align}
 \frac{dk^z}{d\lambda} &=-e E k^t, &
\frac{dk^t}{d\lambda} &=-e E k^z \label{tz}\\
 \frac{dk^x}{d\lambda} &=+e Bk^y, &
 \frac{dk^y}{d\lambda} &=-e Bk^x. \label{xy}
\end{align}
These equations are identical to the ones analysed by Ori \cite{Ori}. In the following  we  make only brief comments about the solutions to these equations, referring the reader to that reference for a more detailed discussion.

We are mostly interested in the motion of a massless particle in the $(t,z)$ plane. The solution for equations \eqref{tz} is:
\bea
 k^t &=&a_1 e^{e E \lambda}+a_2 e^{-e E \lambda}, \label{ktsol} \\
 k^z&=&a_1 e^{e E \lambda}-a_2 e^{-e E \lambda}, \label{kzsol}
\eea
where $a_1$ and $a_2$ are integration constants. From this solution we can see that there are two ways in which a vanishing point can appear,
\begin{align}
  &  (i) & & \hspace{-4cm}a_1=0, \qquad \mbox{and} \qquad \lambda\rightarrow \infty,  \\
  & (ii) & & \hspace{-4cm} a_2=0, \qquad  \mbox{and} \qquad \lambda\rightarrow -\infty.
\end{align}
Once a vanishing point is reached, the continuation beyond is also determined by these two cases. Integrating equations \eqref{ktsol}--\eqref{kzsol} we get
\bea
 t(\lambda) &= & \frac{1}{eE}(a_1 e^{eE\lambda}-a_2 e^{-e E \lambda})+t_c, \\
 z(\lambda)&=&\frac{1}{eE}(a_1 e^{eE\lambda}+a_2 e^{-e E \lambda})+z_c.
\eea
From these expressions we see that the vanishing point is located at finite  spacetime point with coordinates $(t = t_c, z=z_c)$.

Now we are in position to discuss the motion of the orbit beyond the vanishing point. For concreteness we assume that the Lorentz parameter  increases as the vanishing point is approached, i.e, case $(i)$ applies. Under this assumption the orbit approaching the vanishing point $(t = t_c, z=z_c)$ has parameterisation
\begin{align}
 t(\lambda) &=  -\frac{a_2}{eE} e^{-q E \lambda}+t_c, &
 z(\lambda)&=+\frac{a_2}{eE} e^{-q E \lambda}+z_c.
\end{align}
In particular, it has  $\frac{dz}{dt} = -1$, i.e., the $z$ coordinate decreases as the orbit approaches $z=z_c$.
Once the vanishing point is reached either $\lambda$ decreases from $+ \infty$ or it increases from $-\infty$ depending on whether  case $(i)$ or   case $(ii)$ is realised.
 If beyond the vanishing point case $(i)$ is realised then $\frac{dz}{dt} = -1$, i.e., it maintains its sign. Therefore case $(i)$ is the non-bouncing continuation.  On the other hand, in case $(ii)$  $\frac{dz}{dt} = +1$, i.e., it changes  sign across the vanishing point. Therefore case $(ii)$ is the bouncing continuation.

To see which continuation is physical, one can do a perturbative analysis, analogous to the one performed in the previous section, by introducing a small mass parameter and studying timelike Lorentz orbits in flat spacetime with constant electromagnetic field strength. Such an analysis is performed in detail in reference \cite{Ori}. Indeed, one finds that the bouncing continuation is the correct continuation beyond the vanishing point. That analysis applies as it is to the higher-dimensional case.

\section{Charged Vaidya solution: bouncing continuation}
\label{sec:CVS}
In this section we apply the results of the previous sections to get a new interpretation of the charged Vaidya solution.
\subsection{Interpretation}
\label{sec:CVS_interpretation}

Recall that the charged Vaidya solution is supported by the external matter stress tensor \eqref{external_matterST}. The external matter stress tensor  is indeed of the form (cf.~\eqref{matterST}),
\be
T^{\rom{(m)}}_{ab} = \rho k_a k_b,
\ee
however, the normalisation of neither $\rho$ nor that of the null vector $k^a$ is fixed by this expression alone. The extra physics input required to fix these normalisations is provided by the continuity equation \eqref{continuity}.
Assuming $k_a \propto - \partial_a v$, i.e.,  $k^a = S(v,r) \delta^a_r$ implies $J^a = \rho k^a =  \rho(v,r) S(v,r) \delta^a_r.$ Conservation of $J^a$ fixes its form to be
\be
J^a = -\frac{ \sqrt{2(d-2)(d-3)}}{16\pi} \frac{\phi(v)}{r^{d-2}}\delta^a_r, \label{matter}
\ee
where $\phi(v)$ is the flux of particles at retarded time $v$. A convenient numerical factor  is taken out from definition of $\phi(v)$ in equation \eqref{matter} . For the electromagnetic field \eqref{vector},  Maxwell equation \eqref{Maxwell_eq} gives the form of electric current to be
\be
(J^\rom{(e)})^{a} =-\frac{ \sqrt{2(d-2)(d-3)}}{16\pi}\frac{\dot q (v)}{r^{d-2}} \delta^a_r. \label{em}
\ee
Comparing \eqref{matter} and \eqref{em} we find  charge per particle,
\be
e(v) = \frac{\dot q (v)}{\phi(v)}.
\ee
In principle there is nothing wrong in taking the charge per particle $e(v)$ also a function of the retarded time $v$. However, if one imagines the incoming particles to be of the same type at all times, then $e(v)$ is simply a constant.

The expressions for the external energy momentum tensor \eqref{external_matterST} and matter current \eqref{matter} fix both $S(v,r)$ and $\rho(v,r)$,
\bea
S(v,r) &= &  -\sqrt{\frac{2(d-2)}{d-3}}\frac{1}{\phi(v)}\left(\dot m - \frac{q \dot q}{r^{d-3}} \right),\\
\rho(v,r) &=& \left(\frac{d-3}{16 \pi}\right) \frac{\phi^2(v)}{r(r^{d-3}\dot m - q \dot q)}.
\eea
From the expression of $S(v,r)$ we immediately see that
\be
k^r(v,r) = - \kappa_\infty \left( 1 -  \frac{r_c^{d-3}}{r^{d-3}} \right),
\label{kenergy}
\ee
where $\kappa_\infty$  is interpreted as the  energy of the infalling particle measured by a static observer at infinity,
\be
\kappa_\infty = - \lim_{r\to \infty} k^r(v,r) = \sqrt{\frac{2(d-2)}{d-3}}\left(\frac{\dot m(v) e(v)}{\dot q(v)}\right),
\ee
and
\bea
r_c^{d-3} =   \frac{q(v) \dot q(v) }{\dot m(v)}  = \sqrt{\frac{2(d-2)}{d-3}} \left(\frac{ q(v) e(v)}{\kappa_\infty}\right).
\label{rc_eq_rvan}
\eea
Comparing \eqref{region}, \eqref{rvan}, and \eqref{rc_eq_rvan} we see that the critical surface is precisely at the place where the vanishing point of the null vector field $k^a$ is located.
Equation \eqref{kenergy} tells us that the kinetic energy of the thin shell at advanced time  $v$ vanishes  at $r=r_c(v)$. The fact that $\rho$  diverges at $r=r_c(v)$ is not of any consequence; it has no direct physical interpretation\footnote{Note that the number density of these particles measured by an observer is $n_\rom{ob}  = J_a u^a_\rom{ob}  = \rho k_a u^a_\rom{ob}$. This quantity is finite throughout. Since $k_v$ vanishes at $r=r_c$ the problematic factor in $\rho$ is cancelled.}. Moreover, one can easily check that the vector $k^a$ is consistent with the Lorentz force equation \eqref{Lorentz_main}. We conclude that the ingoing charged Vaidya solution is only valid till the surface $r=r_c(v)$.

\subsection{Gluing construction}
\label{sec:CVS_gluing}
In this section we match an ingoing charged Vaidya solution $(\mathM_-,g_-)$ to an outgoing charged Vaidya solution $(\mathM_+,g_+)$ on the bouncing surface $\Sigma$, at which external matter stress tensor vanishes and the null fluid switches from ingoing to outgoing.
Note that, this construction is valid only if the surface $\Sigma$ is spacelike.
If $\Sigma$ is a time-like surface, the ingoing null fluid necessarily intersects the outgoing null fluid near the bouncing surface $\Sigma$. In this case the solution in the overlap region has no simple metric representation. Gluing construction does not seem to work, as there is no surface that separates $(\mathM_-,g_-)$ from $(\mathM_+,g_+)$.
On the other hand, if $\Sigma$ is spacelike, then $\left(\mathM_- \backslash \, \Sigma\right)$ and $\left(\mathM_+ \backslash \, \Sigma\right)$ are disjoint and thus $(\mathM_-,g_-)$ and $(\mathM_+,g_+)$ can be represented by ingoing and outgoing charged Vaidya solutions respectively. Then the gluing can be constructed by imposing suitable matching conditions on the bouncing surface $\Sigma= {\mathM_-} \cap {\mathM_+}$.
\begin{figure}
\begin{center}
\begin{tikzpicture}[domain=-2:3,xscale=0.85,yscale=0.85]
\draw [very thick] plot (\x, {0});
\node [left] at (5,0) {$r=r_c(v)$};
\draw [thick]  [domain=-5:5]  plot ({-2}, {\x});
\draw [dashed] [domain=1:3] plot (\x, {\x-1});
\draw [dashed] [domain=1:3] plot (\x, {-\x+1});
\draw [dashed] [domain=0:3] plot (\x, {\x});
\draw [dashed] [domain=0:3] plot (\x, {-\x});
\draw [dashed] [domain=-1:3] plot (\x, {\x+1});
\draw [dashed] [domain=-1:3] plot (\x, {-\x-1});
\draw [dashed] [domain=-2:3] plot (\x, {\x+2});
\draw [dashed] [domain=-2:3] plot (\x, {-\x-2});
\draw [->]  (0,-2) -- (0,-2.5);
\draw [->]  (0,-2) -- (0.5,-1.5);
\node at (0,-2.75) {$dr$};
\node at (0.65,-1.2) {$dv$};
\draw [->]  (0,2) -- (0.5,1.5);
\draw [->]  (0,2) -- (0,1);
\node at (0,0.7) {$dr$};
\node at (0.7,1.4) {$du$};
\end{tikzpicture}
\captionof{figure}{{\sffamily Ingoing and outgoing charged Vaidya solutions are matched on the spacelike surface $r=r_c(v)$. When the surface $r=r_c(v)$ is not spacelike a simple construction of this type is not possible. Note the direction of the $u$ coordinate.}}
\label{tikz}
\end{center}
\end{figure}
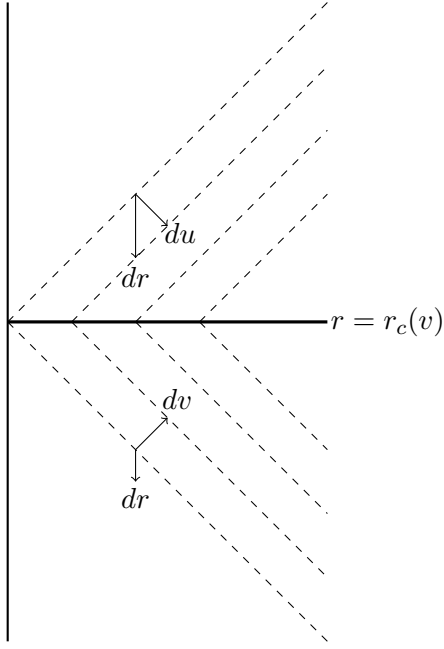

 In particular, we are gluing the spacetime $\mathM_-$ with metric $g_-$ given by ingoing charged Vaidya solution
\be
ds^2 = - f_-(v,r)dv^2 + 2 \, dv \, dr  + r^2 d\Omega^2_{d-2},
\ee
and $\mathM_+$ with metric $g_+$ given by  outgoing charged Vaidya solution\footnote{In this line element, coordinate $u$ is thought of as minus the retarded null coordinate.}
\be
ds^2 = - f_+(u,r)du^2 + 2 \, du \, dr  + r^2 d\Omega^2_{d-2},
\ee
on the surface $\Sigma$, where we have identified radial coordinate $r$ as well as all angular coordinates across $\Sigma$, cf.~figure \ref{tikz}. The function $f_-$ and $f_+$ are given by,
\be
f_-(v,r) = 1- \frac{2 \, m_-(v)}{r^{d-3}} +  \frac{q_-^2(v)}{r^{2(d-3)}} + \frac{r^2}{l^2},
\ee
and
\be
f_+(u,r) = 1- \frac{2 \, m_+(u)}{r^{d-3}} +  \frac{q_+^2(u)}{r^{2(d-3)}} + \frac{r^2}{l^2},
\ee
respectively; $m_-$ and $q_-$ are monotonically increasing positive functions of $v$. The bouncing surface $\Sigma$ is represented in $g_-$ and $g_+$ by,
\bea
r^{d-3} & =& r_{c_-}^{d-3}(v), \\
r^{d-3}  &=& r_{c_+}^{d-3}(u),
\eea
respectively, where,
\be
r_{c_\pm}^{d-3} \equiv \frac{q_\pm \, \dot{q}_\pm}{\dot{m}_\pm}.
\ee
For $r_{c_+}$ the dot denotes derivative with respect to the $u$ coordinate and for $r_{c_-}$ it denotes derivative with respect to the $v$ coordinate.

The preliminary junction condition is then the agreement of the induced metric on $\Sigma$,
\be
\label{pjc}
\left[- \tf_{-}(v) + 2 \,\dot{r}_{c_-}  \right]  \, dv^2  + r^2 d\Omega^2_{d-2}= \left[- \tf_{+}(u) + 2 \,\dot{r}_{c_+} \right]  \, du^2 +   r^2 d\Omega^2_{d-2},
\ee
where, $\tf_{-}(v) \equiv f_-(v,r=r_{c_-}(v))$ and $\tf_{+}(u) \equiv f_+(u,r=r_{c_+}(u))$. Then gluing amounts to finding a function,
\be
u = \psi(v),
\ee
defined on the surface $\Sigma$, such that the preliminary junction condition \eqref{pjc} holds and the extrinsic curvature remains continuous\footnote{Since $\Sigma$ is spacelike, it cannot support  matter or charge density.}$^,$\footnote{These statements are based on comments in \cite{Ori}. In view of claims in
\cite{Booth:2015kxa}, a more detailed study of this issue is required. We leave this for future work.} across $\Sigma$.  Identifying
\begin{align}
 m_+(u) &= m_-(v), & q_+(u) &= q_-(v),
\end{align}
on $ \Sigma$ leads to $r_{c_+}(u) = r_{c_-}(v).$ Inserting this into \eqref{pjc} we get,
\be
\dot{\psi}^2 -  \frac{2 \, \dot{r}_{c_-}}{\tf_-} \, \dot{\psi} + \frac{2 \, \dot{r}_{c_-}}{\tf_-}-1 =0.
\ee
The solutions are,
\begin{align}
\dot{\psi} &=1,  & \dot{\psi} &=  \frac{2 \,\dot{r}_{c_-}}{\tf_-} -1.
\end{align}
The first solution, $\dot{\psi} =1$, corresponds to matching to the non-bouncing continuation and is thus discarded. Choosing $\psi(0)=0$ we obtain a unique matching given by,
\be
\psi(v) = \int_0^v \, \left( \frac{2 \,\dot{r}_{c_-}}{\tf_-} -1\right) dv'.
\ee
From discussion of sections \ref{sec:spherical_symmetry} and \ref{sec:CVS_interpretation}, it follows that after the bounce the null vector field $k^a$ remains future directed, and the null and weak energy conditions are satisfied \cite{Ori}.

\subsection{$q(v)= \nu \  m(v)$}
\label{sec:CVS_qm}
Let us now consider in asymptotically flat settings, a  case in which the mass and the charge functions are proportional to each other, i.e., charge to mass ratio is constant as a function of $v$,
\be
q_-(v) = \nu \, m_-(v),
\ee
where $\nu \in [0,1)$ is a free parameter.  A slightly more general situation is commented on in appendix \ref{app:power}. With this choice of charge distribution the outer and inner apparent horizons are at
\be
r^{d-3}_\pm(v) = \left( 1 \pm \sqrt{1-\nu^2} \right) m_-(v),
\ee
while the radial-coordinate $r$ on the bouncing surface $\Sigma$ depends on $v$ through,
\be
r^{d-3}_{c_-}(v) = \nu^2 \, m_-(v).
\ee
Clearly, the bouncing surface lies in between the inner and outer apparent horizon. The norm of the normal to the surfaces
$S_\lambda \equiv {r^{d-3} - r^{d-3}_{c_-} =\lambda}
$ is
\bea
\left(n_\lambda\right)^a \left(n_\lambda\right)_a &=&  g^{ab} \partial_a S_\lambda \partial_b  S_\lambda \\
 &=& g^{vv} \dot{S_\lambda}^2 + 2 g^{vr} \dot{S_\lambda} S_\lambda' + g^{rr} {S_\lambda'}^2  \\
&=& \left( 2 \dot{S_\lambda}  + f S_\lambda' \right) S_\lambda' \\
& =& (d-3)\left(-2 \nu^2 \dot m_-(v)  + f (d-3) r^{d-4} \right)  r^{d-4}. \label{norm}
\eea
We noted above that the surface $r=r_{c_-}(v)$ lies in between the inner and outer apparent horizons. Therefore, the expression in brackets in equation \eqref{norm} is negative (assuming $\dot m _-(v) > 0$), i.e., the normal to the surface $r=r_{c_-}(v)$ is timelike,
\be
\left(n_\lambda\right)^a \left(n_\lambda\right)_a \big{|}_{\lambda =0}  < 0.
\ee
We conclude that the bouncing surface is spacelike.

However, since the inner apparent horizon lies in the future of this surface, its location must be given in terms of the $u$ coordinate. For a simple situation we provide such an expression now. The matching condition reads,
\be
\psi(v) = - \left( v+\frac{2 \, \nu^4}{1-\nu^2}  m_-(v) \right).
\ee
For concreteness let us consider $m_-(v)$ linear in $v$\footnote{ In four-dimensions this is called the self-similar collapse \cite{Bojowald} as there exists a conformal Killing vector in this situation.  See \cite{Ghosh:2001pv} for generalisation to higher-dimensions. }
\be
m_-(v) = \mu \, v, \qquad 0  \le v \le v_0,
\ee
where $\mu$ is a parameter and  $v_0$ is the width of the shell. Then the ingoing and outgoing coordinates are related on $\Sigma$ by,
\be
u = -\left( 1+\frac{2 \, \nu^4 \, \mu}{1-\nu^2}  \right) \, v.
\ee
The mass function of outgoing null fluid reads,
\be
m_+(u) = - \tldmu \, u,
\ee
where, \be
\tldmu = \mu\left( 1+\frac{2 \, \nu^4 \, \mu}{1-\nu^2}  \right)^{-1} \leq \mu,
\ee
and the charge function $q_+(u) = \nu \, m_+(u)$ remains proportional to mass function by the same factor $\nu$. The inner apparent horizon is at,
\be
r^{d-3}_-(u) = -\left( 1 - \sqrt{1-\nu^2} \right) \tldmu \, u.
\ee

\section{$f(R)$-Maxwell set-up}
\label{sec:fR}
One of the most popular higher-curvature extension of general relativity is the so-called $f(R)$ gravity
\cite{fRReview1, fRReview2, fRReview3}.  A general framework regarding spherically symmetric backgrounds in $f(R)$ theories is well developed \cite{Multamaki:2006zb,Capozziello:2007id, {Bamba:2011sm}}. A large class of vacuum solutions are obtained requiring constant Ricci scalar curvature. For non-vacuum solutions, a requirement to obtain constant Ricci scalar curvature solutions is that the trace of the external stress-tensor should be zero. Since we are interested in charged Vaidya solutions, the requirement of zero trace of the external stress-tensor forces us to work only in four-dimensions.

Fortunately, non-static generalisations, in particular, the four-dimensional charged Vaidya solution with constant Ricci scalar has been worked out \cite{Ghosh:2012zz}. The Lagrangian for this theory is
\be
 L =\left(\frac{1}{16 \pi G}\right) \sqrt{-g} \left[R+f(R)-\frac{1}{4}F^2\right] + L_\rom{matter}.
\ee
The equation of motion for the metric is \cite{Multamaki:2006zb,Capozziello:2007id, Ghosh:2012zz}
\be
 R_{ab}(1+f'(R))-\frac{1}{2}(R+f(R))g_{ab}+\left(g_{ab} \Box - \nabla_a \nabla_b \right)f'(R)=8 \pi T_{ab}^\rom{(em)} + 8 \pi T_{ab}^\rom{(m)},
\ee
where $T_{ab}^\rom{(em)}$ is the stress-tensor of the electromagnetic field \eqref{TEM} and $T^{\rom{(m)}}_{ab}$ is the stress-tensor of the external matter source.
For constant scalar curvature solutions $R=R_0 = $ constant. A charged Vaidya solution with this property takes the form\footnote{In writing some of the expressions below we have fixed certain typos in \cite{Ghosh:2012zz}.}
\be
ds^2=-f(v,r)dv^2+2dvdr+r^2d\Omega_2^2,
\ee
with
\be
 f(v,r)=1-\frac{2m(v)}{r}+\frac{q(v)^2}{r^2(1+f'(R_0))}-\frac{R_0}{12}r^2.
\ee
The only non-vanishing components of $F_{ab}$ are
\be
 F_{rv}=-F_{vr}=\frac{2q(v)}{r^2}.
\ee
The external matter stress tensor required to support the solution is
\be
T_{vv}=  \frac{(1+f'(R_0))}{4 \pi r^2}\left[\dot m (v)-\frac{q(v)\dot q (v)}{(1+f'(R_0))r}\right].
\ee
This stress tensor vanishes at the hypersurface
\be
 r=r_c(v):=\frac{q(v) \dot q(v) }{\dot m(v)(1+f'(R_0))}.
\ee

Following the analysis of section \ref{sec:CVS_interpretation} it is straightforward to show that in this case also, the critical surface is the vanishing surface of the null vector field satisfying the Lorentz force equation. To this end let $k_a=S(v,r) \delta^v_a$.
Once again the continuity equation for the matter and electric current fixes the form of the $S(v,r)$ and $\rho(v,r)$. We find
 \begin{align}
J^a&=-\frac{\phi(v)}{8 \pi r^2}\delta^a_r, &
(J^\rom{(e)})^a &=-\frac{\dot q (v)}{8 \pi r^2}\delta^a_r, &
e(v) &=\frac{(J^\rom{(e)})^a}{J^a}=\frac{\dot q (v)}{\phi(v)}.
\end{align}
Using these expression we have
\begin{align}
 S&=-\frac{2(1+f'(R_0))}{\phi(v)}\left(\dot m (v)-\frac{q(v)\dot q (v)}{(1+f'(R_0))r}\right),\\
 \rho&=\frac{\phi(v)^2}{16 \pi r^2(1+f'(R_0))} \left(\dot m(v)-\frac{q(v) \dot q(v)}{(1+f'(R_0))r}\right)^{-1}.
\end{align}
It can be readily checked that $k^a$ satisfies the Lorentz force equation,
\be
 k^a \nabla_a k^b=e(v)F^b{}_c k^c.
\ee

Qualitatively these expressions are very similar to the expressions in section \ref{sec:CVS_interpretation}. Therefore,  similar considerations apply. The $f(R)$ set-up we have worked with is essentially equivalent to general relativity with a cosmological constant \cite{Ghosh:2012zz}. That is why the analysis of this section is similar to that of the  section \ref{sec:CVS_interpretation}. Nonetheless, we hope that our analysis is a small step that can be viewed as  motivation for a more detailed analysis into other higher-curvature theories.

\section{Discussion}
\label{sec:disc}

We want to emphasise that in the usual interpretation of the charged Vaidya solution null geodesics do enter the $r < r_c$ region; this was precisely the source of confusion in the past literature \cite{Bonnor:1970zz, LakeZannias, SullivanIsrael, Kaminaga:1988pg} (see also \cite{Caceres:2013dma}) . The pathology that there is a region in the spacetime where null and weak energy conditions are violated  is tied to the interpretation of the charged Vaidya solution itself. In the correct interpretation, as first discussed by Ori \cite{Ori}, and further explored in this paper, the pathological region is not there. In this paper we highlighted the fact that the assumption that ``charged photons'' move on null geodesics is inconsistent with the equations of motion. We showed that at $r = r_c$ the Lorentz orbits bounce. The continuation of the Vaidya solution beyond $r = r_c$ surface is unfortunately not a simple matter. Only in the case when $r = r_c$ is a spacelike surface the question can be adequately addressed  by gluing an outgoing charged Vaidya spacetime in the future of the $r = r_c$ surface. When $r = r_c$ surface is timelike the correct interpretation will involve cross-flowing Vaidya solutions, which to the best of our knowledge has no simple metric representation.

In  AdS/CFT correspondence the importance of null energy condition has been realised in several situations. In an attempt to sharpen this connection further, reference \cite{Caceres:2013dma} explored  the strong sub-additivity property of the
holographic entanglement entropy in the
context  of planar charged Vaidya spacetime; a set-up closely related to what we discussed above. They worked on the assumption that the vanishing surface $r=r_c$ separates the entire spacetime in two regions: $(i)$ $r < r_c$ where null energy condition is violated, and  $(ii)$ $r> r_c$ where it is satisfied. They argued that  spacelike minimal surfaces can enter and explore the $r < r_c$ region at least  for certain choices of the mass and charge functions.
Although intentionally focusing on an  unphysical region of a spacetime is not justified from a General Relativity perspective, it can still be useful for other reasons.
Knowing that the charged Vaidya metric for $(0 < r < \infty)$ leads to an unphysical region where the null energy condition is violated,  reference  \cite{Caceres:2013dma}  argued that spacelike minimal surfaces entering the $r < r_c$ region  possibly translates into a violation of strong sub-additivity of entanglement entropy in the dual field theory.

Kaminaga \cite{Kaminaga:1988pg} in 1990 suggested a model for charged black hole evaporation where a negative energy flux of charged null fluid flows into the black hole. The negative energy causes the event horizon to shrink to zero size. Kaminaga studied the model from the classical and semi-classical perspectives. In this model it was found that the inner structure of geometry is very different from that of the Reissner-Nordstr\"om geometry. In this context, the effect of bouncing continuation of charged null orbits has been studied  by Ori and Levin \cite{Levin:1996qt}. They found that the inner structure of black hole suggested by Kaminaga is further drastically  altered when this  effect is taken into account.

In  the mass inflation scenario \cite{Poisson:1990eh}, and in particular in the Ori model of mass inflation \cite{Ori:1991zz}, the charged Vaidya solution plays a very important role. In that context we recall that only the mass is taken to be a function of the advanced time $v$, not the charge, i.e., $q(v)= $ constant. Therefore, the above details do not alter any of the discussions there.

In a different context,  Frolov and Vilkovisky \cite{Frolov:1981mz}  explored effects of higher curvature corrections, as quantum gravity effects, and argued that these effects
cause an incoming null shell to bounce back. Hayward \cite{Hayward:2005gi}  has explored similar ideas as  a model for formation and evaporation of a non-singular black hole. Implications of these ideas for information paradox are  discussed recently by Frolov \cite{Frolov:2014jva}. There are intriguing similarities between our purely classical considerations and ideas put forward in these papers. It will be interesting to explore this link further.

\subsection*{Acknowledgement}
We have benefitted from our discussions with  Swastik Bhattacharya, Naresh Dadhich,  Arnab Kundu, Amos Ori, Sudipta Mukherji, and Sudipta Sarkar.

\appendix
\section{Power law $m(v)$ and $q(v)$ functions}
\label{app:power}
In this appendix we make some  comments on the location of the bouncing surface in asymptotically flat context, having in mind power law $m(v)$ and $q(v)$ functions. In particular, we consider a case where the charge function is related to the mass function as,
\be
q(v)= \nu \, v^s \,  m(v).
\ee
The inner and outer apparent horizons and critical surface are at,
\bea
{r_\pm(v)}^{d-3} &=& \left(1 \pm p \right) m(v), \\
{r_c(v)}^{d-3} &=&  \left(1-p^2 \right)  m(v) \, \theta(v),
\eea
where,
\begin{align}
p(v) &= \sqrt{1- \nu^2 \, v^{2s}}, &
 \theta(v) &=   1 + \frac{s}{v} \frac{m(v)}{\dot m(v)}.
\end{align}
The metric function $\tf$ on the critical surface reads,
\be
\tf(v) = \frac{\left(1-\theta \right)^2 - p^2 \theta^2 }{\left(1-p^2\right) \theta^2}.
\ee
Since $\tf \sim v^{-2 s}$ as $v \to 0$ for $s > 0$, the metric at the critical surface is singular. On the other hand for $s<0$ there exist a $v=v_c$ given by, $v_c^{|s|} = \nu$  below which $r_\pm$ are not real.

\end{document}